\documentclass[amsmath, amssymb, breqn, aps, prd, superscriptaddress, twocolumn, notitlepage, longbibliography]{revtex4-2}

\usepackage[english]{babel}
\usepackage[utf8]{inputenc}
\usepackage{amsthm}
\usepackage{mathtools}
\usepackage{physics}
\usepackage{xcolor}
\usepackage{graphicx}
\usepackage{adjustbox}
\usepackage{placeins}
\usepackage[T1]{fontenc}
\usepackage{lipsum}
\usepackage{csquotes}
\usepackage{float}
\usepackage{units}
\usepackage{tabularx}
\usepackage{booktabs}
\usepackage{bm}
\usepackage{natbib}
\usepackage[colorinlistoftodos, color=green!40, prependcaption]{todonotes}
\usepackage[pdftex, pdftitle={Article}, pdfauthor={Author}]{hyperref} 

\begin{document}
\title{Prototype Superfluid Gravitational Wave Detector}
\author{V. Vadakkumbatt}
    \affiliation{Department of Physics, University of Alberta, Edmonton, AB T6G 2E9, Canada}
\author{M. Hirschel}
    \affiliation{Department of Physics, University of Alberta, Edmonton, AB T6G 2E9, Canada}
\author{J. Manley}
    \affiliation{Department of Electrical and Computer Engineering, University of Delaware, Newark, DE 19716, USA}
\author{T.J. Clark}
    \affiliation{Department of Physics, University of Alberta, Edmonton, AB T6G 2E9, Canada}
\author{S. Singh}
    \affiliation{Department of Electrical and Computer Engineering, University of Delaware, Newark, DE 19716, USA}
\author{J.P. Davis}
    \affiliation{Department of Physics, University of Alberta, Edmonton, AB T6G 2E9, Canada}
    
\begin{abstract}
    We study a cross-shaped cavity filled with superfluid $^4$He as a prototype resonant-mass gravitational wave detector. Using a membrane and a re-entrant microwave cavity as a sensitive optomechanical transducer, we were able to observe the thermally excited high-$Q$ acoustic modes of the helium at $\unit[20]{mK}$ temperature and achieved a strain sensitivity of $\unit[8 \times 10^{-19}]{Hz^{-1/2}}$ to gravitational waves. To facilitate the broadband detection of continuous gravitational waves, we tune the kilohertz-scale mechanical resonance frequencies up to $\unit[173]{Hz/bar}$ by pressurizing the helium.  With reasonable improvements, this architecture will enable the search for GWs in the $\unit[1-30]{kHz}$ range, relevant for a number of astrophysical sources both within and beyond the Standard Model.
\end{abstract}

%\date{\today} % Leave empty to omit a date
\maketitle

\section{Introduction}

Following the breakthrough observation of gravitational waves (GW) with LIGO \cite{Abbott2016a}, and the subsequent observation of numerous additional GW sources \cite{Abbott2016b,Abbott2017a,Krolak2021}, gravitational astronomy is set to become an important facet of multi-messenger astronomy \cite{Branchesi2016,Meszaros2019}. Similar to electromagnetic waves, the future of GW astronomy also demands extending the telescope window to other frequencies. While there has been enormous progress on the low frequency front with pulsar timing arrays~\cite{Arzoumanian2018}, LISA~\cite{Baker2019}, and atom interferometer detectors~\cite{Abe2021}, the high frequency region is often overlooked due to lack of known astronomical GW sources. With over 95\% of the mass-energy content of the Universe invoking beyond the Standard Model physics, it is imperative to extend GW searches to higher frequencies where several interesting sources might lie. Any detection in this window would provide unique insight into the composition of our Universe, or its cosmological evolution. Compact resonant-mass GW detectors, with a small footprint often enabling frequency and geometric tunability, are well suited to search for continuous GW sources in the kHz-GHz window.

One promising candidate for such detectors are high-quality acoustic modes in low-loss superfluid helium, which could have sufficient sensitivity to detect small strains induced by gravitational waves when combined with low-noise transduction \cite{DeLorenzo2014,DeLorenzo2017,Singh2017}. Recently, a prototype of a superfluid helium based detector with microwave cavity readout was experimentally demonstrated in a cylindrical geometry \cite{DeLorenzo2014,DeLorenzo2017}. Despite the very high mechanical and microwave $Q$'s achieved, this architecture was unable to resolve thermally driven mechanical motion. Here, we study the high-$Q$ mechanical modes of a cross-shaped superfluid helium resonator, read out with a membrane and re-entrant microwave cavity as a sensitive optomechanical parametric transducer, which allows us to observe the thermally-limited motion of a $\unit[4.28]{g}$ resonator down to $\unit[20]{mK}$. The high mechanical $Q$ allows us to realize GW strain sensitivities down to $\unit[8 \times 10^{-19}]{Hz^{-1/2}}$, while the cross shape yields gravitational cross sections as large as $\unit[47]{\%}$ of the physical area. Notably, because of the dependence of the helium speed of sound on pressure, we are able to continuously tune the acoustic frequency of the resonant-mass detector, up to $\unit[173]{Hz/bar}$, something that cannot be done in contemporary GW observatories and is a major limitation of solid Weber-bar antennas \cite{Michelson1984}.

A tunable, narrowband GW detector made of superfluid helium as the one demonstrated here would be sensitive to GWs in the 1-30 kHz range. This system complements other proposed resonant-mass detectors for high frequency GWs such as levitated dielectrics \cite{Arvanitaki2013} and bulk acoustic wave resonators \cite{Goryachev2014}. In this paper, we discuss sources of GWs in the kHz regime followed by the demonstration of thermal motion-limited readout and pressure-induced tunability of the prototype superfluid GW detector. %made of 4.28 grams of helium coupled to a re-entrant microwave cavity. 
Finally, we comment on the feasibility of using superfluid helium optomechanical systems as high frequency GW detectors.

\section{GW sources in the kHz regime}

\begin{figure}[t]
	\centering
	\includegraphics{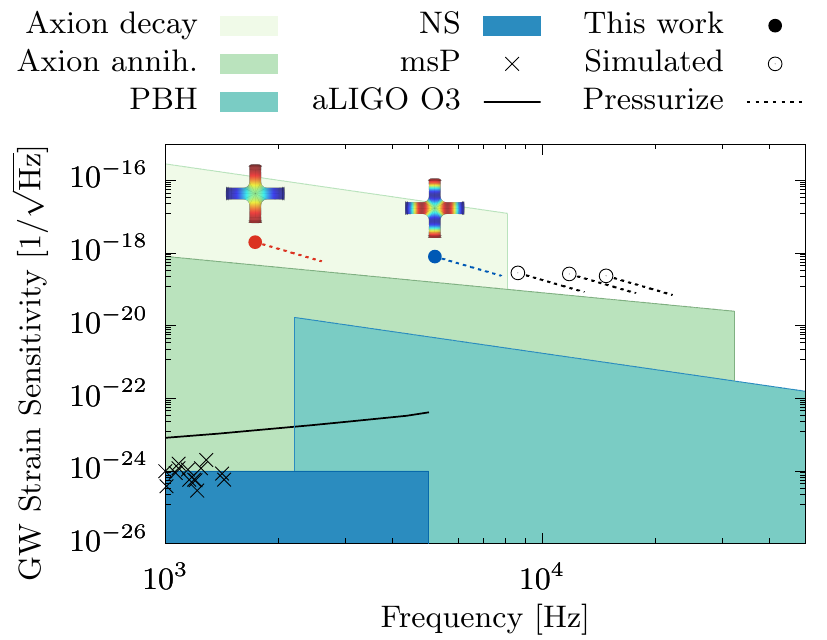}
	\caption{Estimations of the required sensitivity to detect GWs from various sources: superradiance from axion decay and axion annihilation, binary primordial black hole (PBH) mergers, binary neutron star (NS) post-merger remnants, and millisecond pulsars (msP). The measured sensitivities for the [1,-1] and [2,-2] modes of the prototype from this work are included, as well as the simulated minimum sensitivities for three higher-order harmonics \cite{Singh2017}. The dotted lines show the expected performance when the frequencies of these modes are tuned through pressurization. Also included is Advanced LIGO's noise budget from its third observing run (aLIGO O3) for comparison \cite{Buikema2020}.}
	\label{fig:GWparameterspacePlot}
	\vspace{-5mm}
\end{figure}

In this section, we briefly summarize some potential sources of GWs above $1$ kHz, such as mergers of neutron stars (NS) and primordial black holes (PBH), millisecond pulsars (msP), and black hole (BH) superradiance \cite{Aggarwal2020}. Mergers of binary NSs and PBHs are expected to produce transient GWs, while msPs and BH superradiance will produce monochromatic GWs, which makes them better candidates for a GW search using a resonant detector. See Ref.~\cite{Aggarwal2020} for a more comprehensive treatment of high-frequency GWs. Estimations of the strain sensitivities required to detect GWs from each source are shown in Fig.~\ref{fig:GWparameterspacePlot}. 

Neutron star binaries in the late inspiral phase emit detectable transient GWs, which have been observed by LIGO/Virgo \cite{Abbott2017d}. If the NS binary does not immediately collapse to form a BH, a post-merger remnant will continue to emit weaker GWs for up to $\mathcal{O}\left(100\rm ms\right)$ due to its oscillation and rotation \cite{Bernuzzi2015}. The post-merger signal contains information about the remnant's high-density equation of state \cite{Bauswein2016,Ackley2020}, but has not yet been detected \cite{Abbott2019}. In Fig.~\ref{fig:GWparameterspacePlot} we use $\lesssim 10^{-24}$ Hz$^{-1/2}$ as an estimate \cite{Torres2019} of the strain sensitivity required to begin detecting post-merger GW emission from binary NSs over a frequency range of $1-5$ kHz.

Transient GWs at kHz frequencies can also be sourced by mergers of sub-solar mass BHs, which would likely have a primordial origin \cite{Aggarwal2020}. Figure \ref{fig:GWparameterspacePlot} estimates the required sensitivity to detect binary PBH mergers at a distance of 10 kpc, assuming the PBHs to have equal mass of $M_\text{BH}$ (using equations (19) and (20) from Ref.~\cite{Aggarwal2020}). The low-frequency cutoff at $\approx 2$ kHz comes from an assumed upper-bound of $M_\text{BH} \leq M_\odot$ on the individual PBH mass \cite{Aggarwal2020}.  

Non-axisymmetric neutron stars in our galaxy are the most well-studied source of continuous gravitational waves. Data from Advanced LIGO has already placed constraints on stellar ellipticity for 222 pulsars with rotation frequencies above 10 Hz \cite{Abbott2017c}. Pulsars would emit continuous GWs at twice their rotation frequency, $2f_p$ \footnote{There are mechanisms for GW emission at the pulsar rotation frequency, as well \cite{Abbott2017c}. However, they are not considered in this work due to their low frequency.}. In Fig.~\ref{fig:GWparameterspacePlot}, we plot the spin-down limit on gravitational strain for the 14 pulsars mentioned in Table \ref{tab:MSPtable} with rotation frequencies above 500 Hz (gravitational wave frequency $2f_p$) that were considered in the last two LIGO surveys \cite{Abbott2017c, Aasi2014}. The spin-down limit is obtained by equating the power loss from the observed spin-down of the pulsar's rotational frequency being due to emission of gravitational waves \cite{Riles2017}. Monitoring targeted pulsars electromagnetically would allow for coherent integration of the GW signal over long times by a narrowband detector, such as the one demonstrated here. 

Another potential source of monochromatic GWs are boson clouds that may form around rapidly rotating BHs through superradiance.  This phenomenon assumes the existence of beyond the Standard Model massive bosons such as axions \cite{Arvanitaki2010}, dark photons \cite{Siemonsen2020}, or tensor fields \cite{Brito2020}, which have emerged as dark matter candidates. If the boson's Compton wavelength is comparable to the BH size (in terms of the Compton frequency, $f_\text{c} \approx c^3 / 4 \pi G M_\text{BH}$), they may populate bound states around a BH, forming a ``gravitational atom'' \cite{Arvanitaki2013}. The boson cloud will emit monochromatic GWs as the particles transition to lower energy levels \cite{Arvanitaki2015}, annihilate with each other \cite{Arvanitaki2013}, or decay into gravitons \cite{Sun2020}. Figure \ref{fig:GWparameterspacePlot} estimates the required sensitivity to detect GWs for axion annihilation and decay processes assuming a distance of 10 kpc. The GW amplitude from axion annihilation is estimated using equation (7) in Ref.~\cite{Arvanitaki2013}, where the GW frequency is twice the Compton frequency. The GW amplitude from axion decay is estimated using equation (34) in Ref.~\cite{Sun2020}, where the GW frequency is half the Compton frequency. For both cases we impose a lower bound $M_\text{BH}\geq M_\odot$ on the BH mass. 

\begin{table}[b]
    \centering
    \caption{Millisecond pulsars with frequency greater than 500 Hz considered in the last two continuous GW surveys by LIGO+VIRGO collaboration \cite{Aasi2014,Abbott2017c}: $f_p$ is the rotational frequency, $f_{GW} =2 f_p$ is the frequency of gravitational waves, $d$ is the distance to the pulsar in kiloparsecs, $h_{sd}$ is the spin-down strain limit.}
    \label{tab:MSPtable}
    \vspace{0.25cm}
	\begin{tabularx}{\columnwidth}{lX|XcXcXcXc}
		Pulsar &&& $f_p$ [Hz] && $f_{GW}$ [Hz] && $d$ [kpc] && $h_{sd}$  \\
		\cmidrule{1-10}
		J0034-0534 &&& 532.7 && 1065.4 && 1.35 && $8.9 \times 10^{-28}$  \\
		J0952-0607 &&& 707.3 && 1414.6 && 1.74 && $8.5 \times 10^{-28} $ \\
	    J0955-61 &&& 500.2 && 1000.4 && 2.17 && $9.9 \times 10^{-28} $  \\
		J1301+0833 &&& 542.4 && 1084.8 && 1.23 && $1.6 \times 10^{-27} $ \\
		J1747-4036 &&& 607.7 && 1215.4 && 7.15 && $2.9 \times 10^{-28} $  \\
		\cmidrule{1-10}
		J1748-2446O &&& 596.4 && 1192.8 && 5.53 && $5.8 \times 10^{-28} $  \\
		J1748-2446P &&& 578.5 && 1157 && 5.53 && $5.8 \times 10^{-28} $  \\
		J1748-2446ad &&& 716.4 && 1432.7 && 5.53 && $5.8 \times 10^{-28} $  \\
		J1810+1744 &&& 601.4 && 1202.8 && 2.36 && $5.6 \times 10^{-28} $ \\
		J1843-1113 &&& 541.8 && 1083.6 && 1.48 && $1.2\times 10^{-27} $  \\
		\cmidrule{1-10}
		J1902-5105 &&& 573.9 && 1147.8 && 1.65 && $1.1 \times 10^{-27} $ \\
		J1939+2134 &&& 641.9 && 1283.8 && 3.27 && $2.0 \times 10^{-27} $ \\
		J1959+2048 &&& 622.1 && 1244.2 && 1.73 && $1.2 \times 10^{-27} $ \\
		J2052+1218 &&& 503.7 && 1007.4 && 3.92 && $3.8 \times 10^{-28} $ \\
	\end{tabularx}
\end{table}

For monochromatic and long-lived GWs, such as those sourced by msPs and BH superradiance, we assume a measurement time of $\tau_\text{int}=10^6$ seconds, which relaxes the sensitivity requirements ($S_{hh} \propto \tau_\text{int}$). The frequency tunability of our system will allow for Doppler frequency shift corrections due to earth's rotation and orbital motion \cite{Schutz1989}. The small size of the detector enables maximizing the geometric overlap with a GW source at known sky position. Moreover, the demonstrated frequency tunability would allow the same acoustic resonator to look for GW signals from multiple sources.  Along with the expected GW strains shown in Fig.~\ref{fig:GWparameterspacePlot}, we show the demonstrated sensitivities of the prototype detector, and simulated sensitivities for higher order modes, which we describe in further detail below. 

\section{Experimental Design}

\begin{figure}[t]
    \centering
    \includegraphics{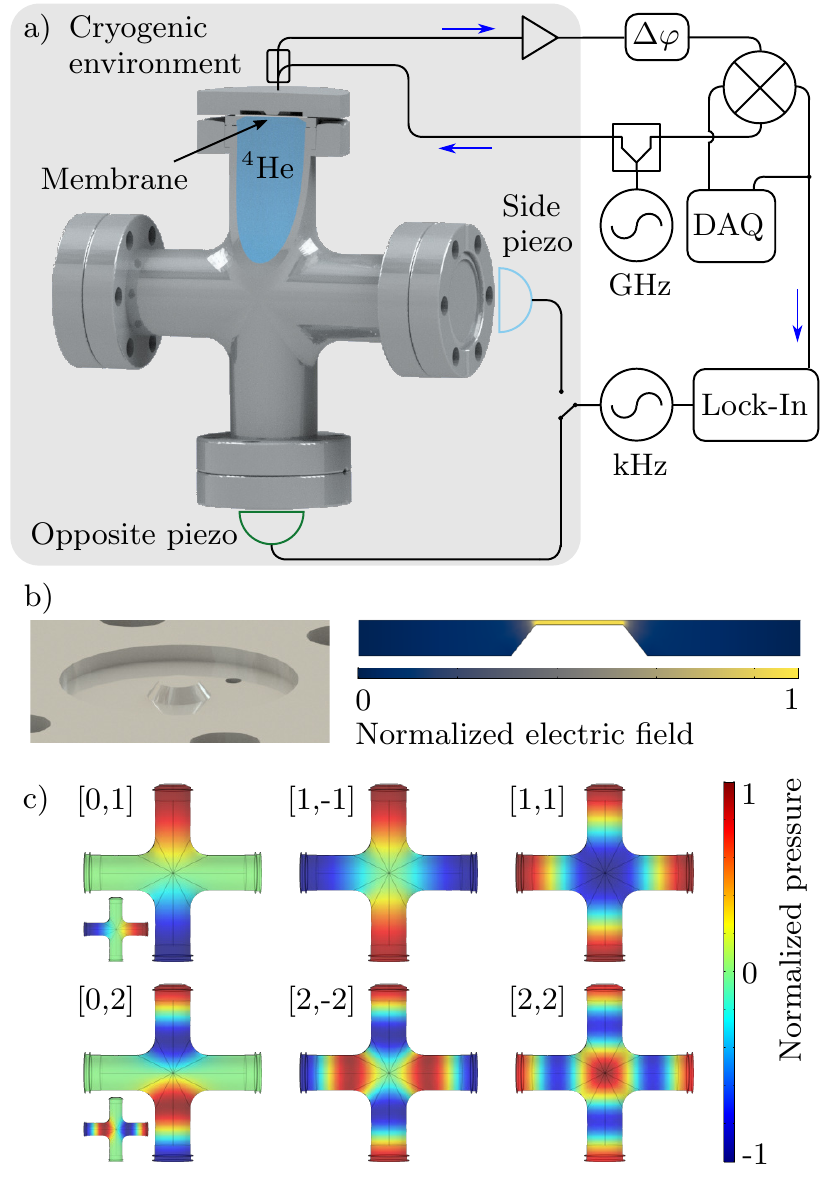}
    \caption{a) Sketch of the experimental setup. The grey box represents the cryogenic environment, with the homodyne detection electronics at room temperature. We can switch between driven ring-down measurements detected with the lock-in amplifier and un-driven thermomechanical measurements. b) Detailed view of the bottom half of the re-entrant microwave cavity (left) showing the re-entrant stub, with electric field amplitude (right). c) Finite element simulations of the normalized pressure fields of the helium acoustic modes coupled to the fundamental membrane flexural mode. The helium modes are labeled according the the relative pressure of the two arms [x,y].}
    \label{fig:SetupSketch}
\end{figure}

Figure \ref{fig:SetupSketch} shows a schematic of the detector architecture. We used a commercial 1.33" conflat-flange 4-way-cross made of stainless steel, as this provides a straightforward superfluid-leak-tight sample cell. Three arms of the cross were capped with standard conflat flanges, while the fourth was hermetically sealed with a circular aluminum membrane of $\unit[250]{\mu m}$ thickness affixed to a copper gasket. The outer side of the membrane acted as one face of a cylindrical microwave cavity, similar to previous designs \cite{Clark2018,Potts2020}. The other half of the microwave cavity was machined from 6061 aluminum and incorporates a re-entrant stub (Fig.~\ref{fig:SetupSketch}b). A nonlinear regression of the reflection scattering parameter revealed an internal microwave cavity quality factor of $\approx 1.5 \times 10^4$ \cite{Probst2015}. As the pressure in the helium cell is increased, the aluminum membrane is displaced towards the re-entrant stub and the gap --- of $\approx \unit[100]{\mu m}$ and where most of the electric field is concentrated --- decreases. In a lumped circuit element model, this increases the capacitance, which decreases the resonance frequency of the microwave cavity \cite{Noguchi2016}.  This enables sensitive microwave readout of the pressure fluctuations in the helium cell, with a single-photon single-phonon coupling rate of up to $|g_0| = \unit[2.8 \times 10^{-5}]{Hz}$ \cite{Aspelmeyer2014}. 

Two broadband piezoelectric transducers were epoxied to the conflat flanges opposite and perpendicular to the arm with the microwave cavity. These piezos enabled a controlled excitation of the acoustic modes of the helium, and hence mode identification. The cell fill line was soldered into the center of the cross, at a pressure node for the GW-relevant acoustic modes. To minimise mechanical losses, the whole assembly was hung from the bottom of the dilution fridge via thin wires. A thermal link to the mixing-chamber plate was made via flexible copper braids. The microwave components and helium fill line were well thermalized, such that the experiment had a minimum base temperature of $\unit[16]{mK}$, as measured by a primary nuclear-orientation thermometer.  Outside the fridge, we used a standard homodyne detection setup (outlined in Fig.~\ref{fig:SetupSketch}) to measure the harmonic phase fluctuations caused by the frequency modulation of the microwave cavity \cite{Teufel2011a}.

\section{Characterization}

Figure \ref{fig:SetupSketch}c shows the normalized pressure fields of the first six acoustic modes of the helium-membrane system simulated with finite element modelling. Table \ref{tab:ModeParameters} lists the corresponding effective masses $\mu \equiv \int \rho \bm{u}^2 \, \text{d}V$ (with density $\rho$, unit amplitude displacement field $\bm{u}$ of the helium and membrane normal mode, and volume $V$) and GW effective areas $A_G \equiv 2 \sum q_{ij}^2 / \mu M$ \cite{Hirakawa1976} (with dynamic part of the quadrupole moment $q_{ij} \equiv \int \rho \, ( u_i x_j + x_i u_j - 2/3 \, \delta_{ij} \bm{u} \cdot \bm{x}) \, \text{d}V$). With GW quadrupolar strain in mind, the simulation confirms the expectation that the acoustic modes labeled $[1,-1]$ and $[2,-2]$ feature the largest GW effective area. Therefore, in the following we will focus our analysis on these two modes. Simulations also show that the uncoupled fundamental flexural membrane mode has a frequency of $\unit[12.2]{kHz}$, significantly higher than the acoustic mode frequencies of the helium. Therefore, the normal mode frequencies of the oscillators essentially decouple and the membrane displacement is expected to follow the helium motion instantaneously \cite{Rapagnani1982}.

\begin{table}[b]
    \centering
    \caption{Relevant parameters of the helium-membrane normal modes from finite-element simulations: mechanical frequency $f_\text{m}$, effective mass $\mu$ in units of the geometric mass $M = \unit[4.28]{g}$, and GW effective area $A_G$ in units of the helium cross-sectional area $A = \unit[22.9]{cm^2}$ \cite{Hirakawa1976}. Only the first six modes were considered in the experiment.}
    \vspace{0.25cm}
    \begin{tabularx}{\columnwidth}{l|cccXl|ccc}
    Mode & $f_\text{m}$ [Hz] & $\mu/M$ & $A_G/A$ & & Mode & $f_\text{m}$ [Hz] & $\mu/M$ & $A_G/A$\\
    \cmidrule{1-4}
    \cmidrule{6-9}
    {[0,1]} & 1618 & 0.16 & 0 & & {[2,2]} & 6381 & 0.41 & 0.04\\
    {[1,-1]} & 1769 & 0.19 & 0.47 & & & & &\\
    {[1,1]} & 3179 & 0.47 & 0.21 & & {[3,-3]} & 8619 & 0.27 & 0.07\\
    {[0,2]} & 4692 & 0.26 & 0 & & {[4,-4]} & 11800 & 0.15 & 0.03\\
    {[2,-2]} & 5253 & 0.21 & 0.27 & & {[5,-5]} & 14770 & 0.12 & 0.02\\
    \end{tabularx}
    \label{tab:ModeParameters}
\end{table}

\begin{figure}[t]
    \centering
    \includegraphics{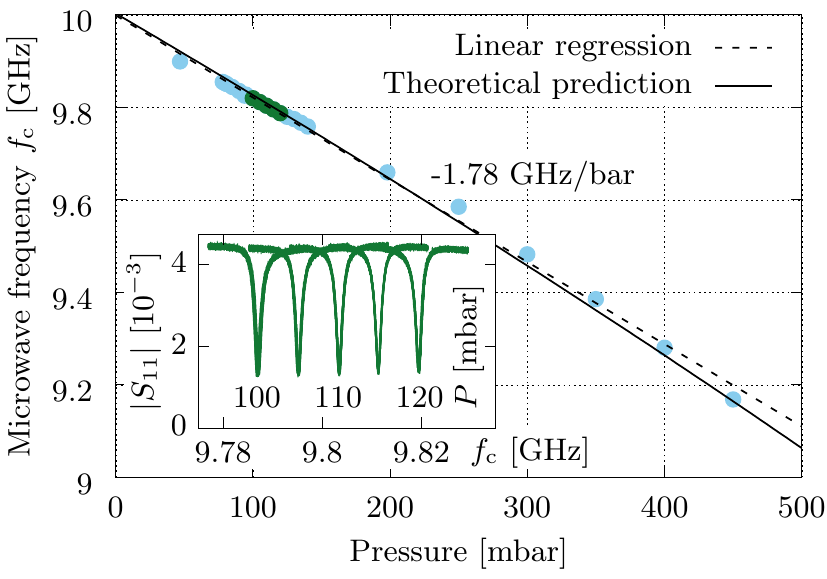}
    \caption{Tunability of the microwave cavity frequency by changing the helium pressure, which allows for calibration of the optomechanical coupling rate. The microwave cavity responds nearly linearly to pressure (dashed line) although the dependence is more accurately described by a nonlinear model \cite{Fujisawa1958} (solid line).  
    The inset shows the microwave reflection scattering parameter $S_{11}$ at five example pressures (dark green points) between $P = 100$ and $\unit[120]{mbar}$.  Each trace is actually $50$ (overlapping) curves, with no discernible difference, demonstrating the stability of the detector.}
    \label{fig:f0vsP_Inset}
\end{figure}

As our detection scheme is based on a highly tunable microwave cavity, it is possible to directly measure the acoustic to microwave optomechanical coupling strength by varying the helium pressure and tracking the resonance frequency of the microwave cavity. After filling the cross with liquid helium and cooling to $\unit[20]{mK}$, the pressure inside the cell was varied from $P = \unit[80]{mbar}$ to $\unit[450]{mbar}$. The microwave cavity resonance frequency $f_\text{c}$ decreases approximately linearly with pressure, as the membrane moves towards the stub. The result is shown in Fig.~\ref{fig:f0vsP_Inset} and a linear regression to the data (dashed line) provides an acoustic to microwave coupling strength of $\partial f_\text{c}/\partial P = \unit[-1.78]{GHz/bar}$. Also shown is a nonlinear single-parameter fit \cite{Fujisawa1958} (solid line), predicting the frequency as function of geometric cavity parameters. Using the zero point pressure fluctuations $\Delta P_0 = \sqrt{h f_\text{m}/\kappa V_\text{eff}}$ of the acoustic mode \cite{DeLorenzo2016} (with the acoustic mode frequency $f_\text{m}$, helium compressibility $\kappa = \unit[1.2 \cdot 10^{-7}]{Pa^{-1}}$, and simulated effective volume $V_\text{eff}$ of the respective pressure mode), the resulting single-photon single-phonon coupling rate is given as $g_0 \equiv (\partial f_\text{c}/\partial P) \, \Delta P_0 = \unit[-1.6 \times 10^{-5}]{Hz}$ for the $[1,-1]$ mode and $g_0 = \unit[-2.8 \times 10^{-5}]{Hz}$ for the $[2,-2]$ mode \cite{Aspelmeyer2014}. These values are three orders of magnitudes larger than the previous GW superfluid detector prototype \cite{DeLorenzo2014}. As a result, even at $\unit[20]{mK}$, the detector sensitivity is limited by thermal noise. 

\begin{figure}[t]
    \centering
    \includegraphics{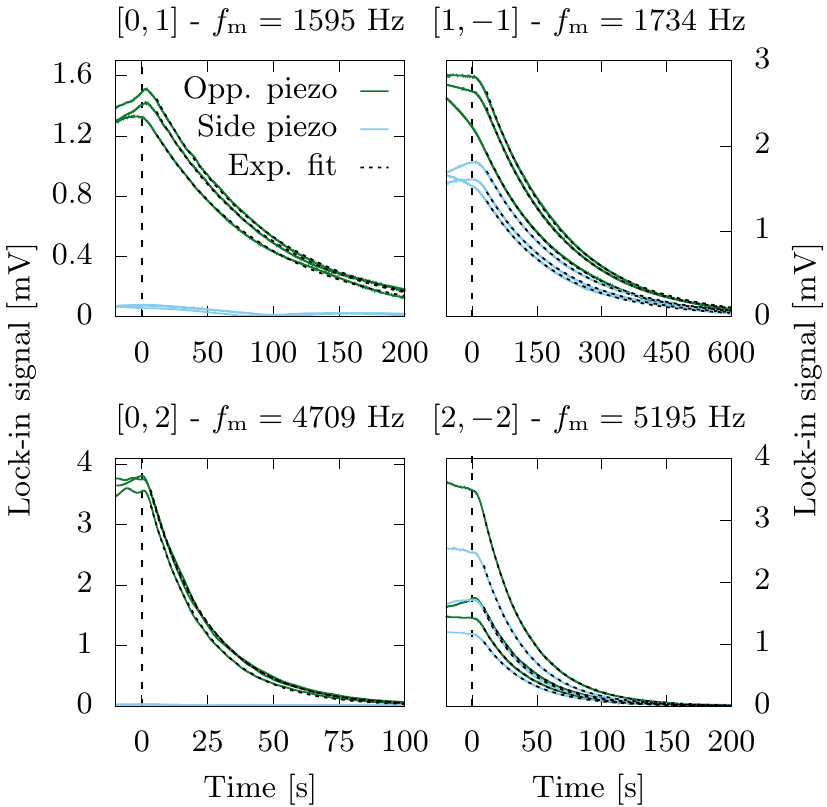}
    \caption{Comparison of ring-down measurements of four normal modes driven by both the opposite (green) and side (blue) piezoelectric actuators. The $[0,1]$ and $[0,2]$ modes can only be excited with the opposite piezo, since the measurable oscillation takes place in the vertical arms shown in Fig.~\ref{fig:SetupSketch}c, confirming mode identification.  Fits to such ring-downs allow unambiguous extraction of mechanical $Q$'s, as discussed in the text and shown for the $[1,-1]$ and $[2,-2]$ modes in Fig.~\ref{fig:QvsT}}
    \label{fig:PiezoComparison}
\end{figure}

To identify the dominant loss mechanisms of the mechanical modes in our detector, the quality factors of the modes were evaluated through ring-down measurements. After excitation, by driving the opposite or side piezo at the mode resonance frequency $f_\text{m}$, the voltage output of the mixer was measured using a lock-in amplifier locked to $f_\text{m}$, yielding a signal decaying as $V(t) = V_0 \, \exp(-t/\tau)$. Then, the $Q$ was determined by fitting the decay time constant $\tau = Q/(\pi f_\text{m})$ of the data. Figure \ref{fig:PiezoComparison} shows exemplary ring-down measurements for four normal modes. The $[0,1]$ and $[0,2]$ modes were not efficiently actuated using the piezoelectric transducer perpendicular to the membrane (the side piezo), which supports mode identification.

\begin{figure}[t]
    \centering
    \includegraphics{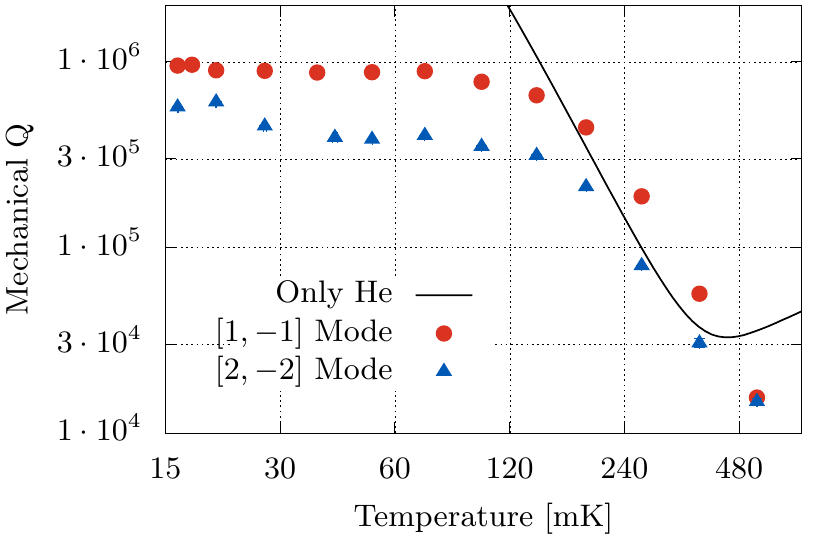}
    \caption{Mechanical quality factors of the $[1,-1]$ and $[2,-2]$ modes versus temperature, as well as the theoretical prediction for three-phonon dissipation in pure $^4$He \cite{Abraham1969}.}
    \label{fig:QvsT}
\end{figure}

The quality factors of the $[1,-1]$ and $[2,-2]$ mode were measured in a temperature range from $16$ to $\unit[500]{mK}$ (see Fig.~\ref{fig:QvsT}). The dominant intrinsic dissipation mechanism in superfluid helium is a three-phonon process \cite{Abraham1969}, which limits the quality factor of the acoustic modes below $\unit[350]{mK}$ with a temperature dependence following $Q \propto T^{-4}$ (solid line in Fig.~\ref{fig:QvsT}). 

\begin{figure}[b]
    \centering
    \includegraphics{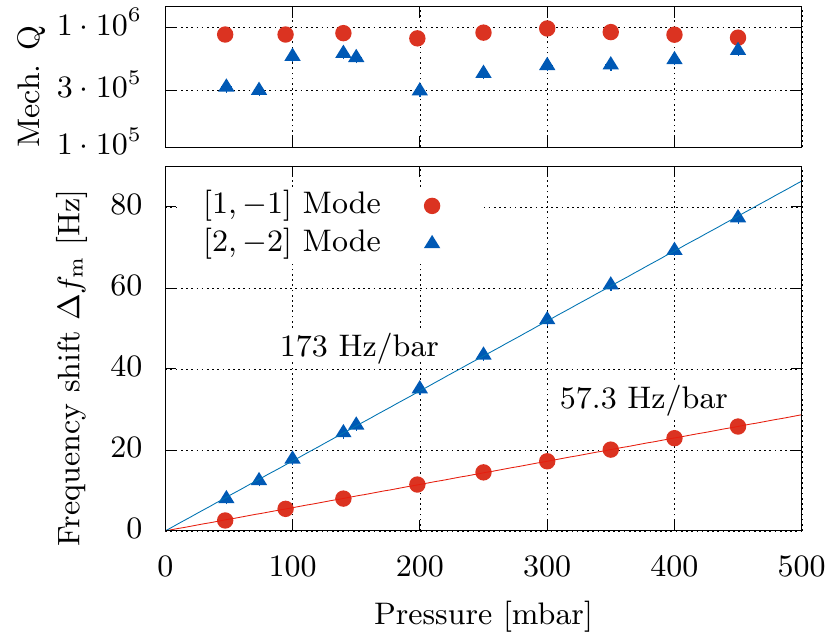}
    \caption{Mechanical quality factor $Q$ and frequency shift $\Delta f_\text{m}$ with linear regressions of the $[1,-1]$ and $[2,-2]$ modes versus pressure. The frequencies at zero pressure are $f_\text{m} = \unit[1726]{Hz}$ and $\unit[5171]{Hz}$, respectively.}
    \label{fig:f0_Q_vsP}
\end{figure}

The measured data agrees with the theory between $150$ and $\unit[350]{mK}$. Below $\unit[150]{mK}$, the quality factors saturate to $Q \approx 10^6$ for the $[1,-1]$ mode and $Q \approx 5 \times 10^5$ for the $[2,-2]$ mode. In this temperature regime, the motion of the aluminium membrane is likely the dominant source of mechanical dissipation limiting the quality factor. If this were the case, one would expect that the losses would be worse for the $[2,-2]$ normal mode, as observed, since its frequency is closer to the uncoupled membrane frequency of $\unit[12.2]{kHz}$. This, however, requires further experimental investigation. Other possible channels of dissipation in our system are $^3$He impurities in the superfluid $^4$He \cite{Kerscher2001}; acoustic losses in the mechanical suspension or thermal connections; as well as helium in the fill line \cite{DeLorenzo2017}.

One of the unique features of superfluid helium based resonant-mass detectors is the tunability of the mechanical frequency via changing the speed of sound through pressurization \cite{Brooks1977,Abraham1970}. This makes a single narrow-band detector sensitive to a broad frequency range, and allows one to track and monitor a GW source for a longer duration by applying frequency corrections \cite{Schutz1989}. We have investigated this feature by varying the pressure and measuring the change of the mechanical frequency. The results for the $[1,-1]$ and $[2,-2]$ mode are shown in Fig.~\ref{fig:f0_Q_vsP} for a pressure range between $50$ and $\unit[450]{mbar}$. The mechanical quality factors (top panel) of the respective modes remain essentially unaffected by the pressure tuning. The frequency shift (bottom panel) is well reproduced by linear regressions revealing a tunability of $\unit[57.3]{Hz/bar}$ for the $[1,-1]$ and $\unit[173]{Hz/bar}$ for the $[2,-2]$ mode.

\section{Thermally Excited Acoustic Modes}

In an ideal scenario, one will have a resonant-mass GW detector that is not limited by so-called technical noise, but instead by intrinsic noise sources, such as shot noise or thermal noise \cite{Michelson1987,Clerk2010,Purdy2013}.  In this case, shot noise can be reduced by increased measurement power, and hence thermal Brownian motion will generally limit the sensitivity of resonant-mass detectors. Therefore, it is relevant to cool such detectors to as low of a temperature as possible, to reduce the thermal noise contribution and improve the GW strain sensitivity.  For liquid $^4$He, this has the added benefit of reducing the mechanical dissipation, as shown in Fig.~\ref{fig:QvsT}.  Here, due to the high acoustic-to-microwave coupling strength, we were able to measure the thermally excited modes of the helium in the cross geometry at a temperature of $\unit[20]{mK}$. 

\begin{figure}[t]
    \centering
    \includegraphics{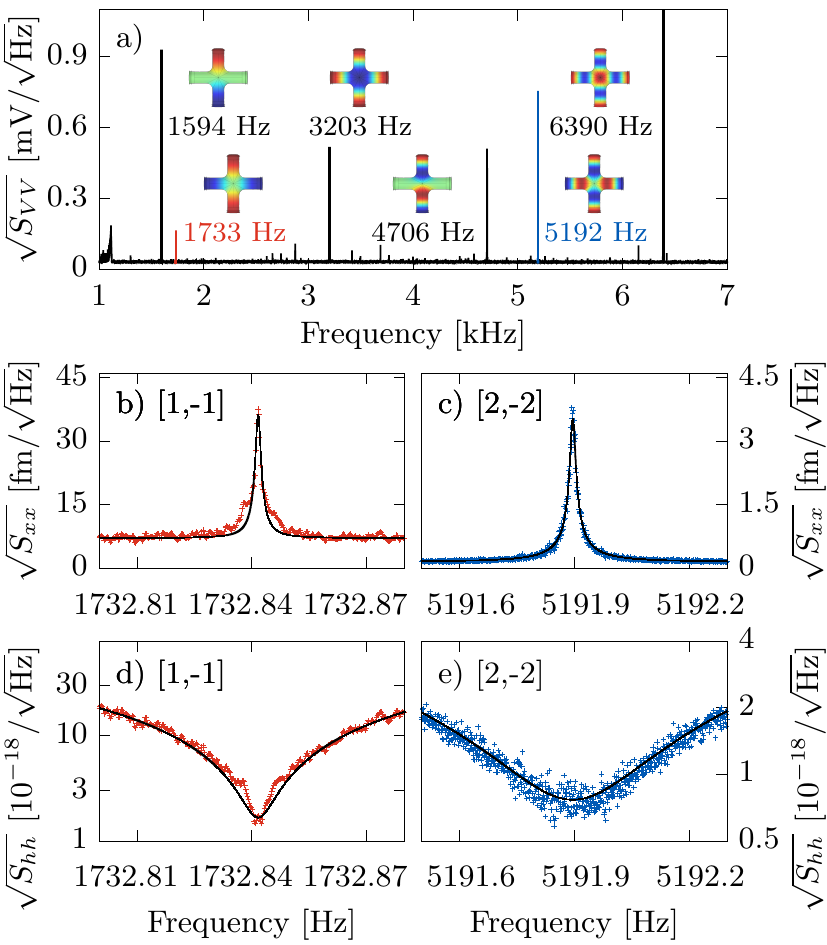}
    \caption{Power spectral densities resulting from thermomechanical motion at a temperature of $\unit[20]{mK}$. a) Voltage spectral densities $\sqrt{S_{VV}}$ over the entire measured frequency range, with simulations of the mode shapes and measured frequencies.  The two GW-relevant modes are accentuated, red for the $[1,-1]$ and blue for the $[2,-2]$ mode, throughout. Calibrated displacement spectra $\sqrt{S_{xx}}$ are shown for the b) $[1,-1]$ mode and c) $[2,-2]$ mode. The resulting GW strain sensitivities $\sqrt{S_{hh}}$ are shown in d) for the $[1,-1]$ mode and e) for the $[2,-2]$ mode.}
    \label{fig:FullPSD_1734Hz_5195Hz}
\end{figure}

To achieve high signal to noise, 71 undriven time series datasets, each having a length of $\unit[1024]{s}$ ($\approx \unit[17]{minutes}$), were independently acquired and averaged together. During data acquisition, small drifts in the pressure resulted in small drifts in the power spectral densities, of $\approx \unit[50]{mHz}$, which were individually compensated before averaging the 71 data sets together.  Note that these pressure fluctuations (on the scale of a mbar) will be minimized in future iterations by introducing a cold valve to seal the sample cell after filling and pressurization \cite{DeLorenzo2017}. The resulting voltage power spectral density $S_{VV}$ is shown in Fig.~\ref{fig:FullPSD_1734Hz_5195Hz}a. The observed resonance frequencies $f_\text{m} = \omega_\text{m}/2 \pi$ are in good agreement with those simulated (see Table \ref{tab:ModeParameters}). 

To calibrate the measurement in terms of a displacement spectral density $S_{xx} \propto S_{VV}$, we followed the method of Hauer et al.~\cite{Hauer2013}, based on the fluctuation-dissipation theorem which predicts a mean squared thermal noise force of 
\begin{equation}
    \label{Eq:ThermalNoiseForce}
    S_{FF}^\text{th} = 4 k_\text{B} T \, \frac{\omega_\text{m} \mu}{Q}
\end{equation}
for each mode, with measured temperature $T = \unit[20]{mK}$ and effective mass $\mu$. Figures \ref{fig:FullPSD_1734Hz_5195Hz}b and c show the calibrated displacement spectra of the $[1,-1]$ and $[2,-2]$ modes, respectively. The readout noise-limited sensitivity of mechanical displacement is $\unit[7.1]{fm ~ Hz^{-1/2}}$ for the $[1,-1]$ mode and $\unit[0.16]{fm ~ Hz^{-1/2}}$ for the $[2,-2]$ mode.  These values are comparable with the state-of-the-art microwave cavity optomechanics \cite{Teufel2011b}.

The sensitivity to strain oscillations induced by continuous GWs, with a strain power spectral density $S_{hh}(\omega)$, can be resonantly enhanced, resulting in a force spectrum of \cite{Singh2017}
\begin{equation}
    \label{Eq:GWForceStrainRelation}
    S_{FF}(\omega) = \frac{1}{40} \, M \mu \omega^4 A_G d \, S_{hh}(\omega),
\end{equation}
acting on the resonant-mass antenna with geometric mass of the helium $M$, GW effective area $A_G$, and directivity function $d(\theta,\varphi,\Psi)$. The directivity function depends on the angular orientation $(\theta,\varphi)$ and polarization $\Psi$ of the GW relative to the detector, and can be maximized to $d_\text{max} = 2.5$ for the $[1,-1]$ and $[2,-2]$ modes \cite{Hirakawa1976}. Inverting Eqn.~\eqref{Eq:GWForceStrainRelation}, and using the force spectral densities $S_{FF} = S_{xx}/|\chi^2(\omega)|$ corresponding to the measured noise spectra (with mechanical susceptibility $\chi(\omega) = [\mu \, (\omega_\text{m}^2 - \omega^2 + i \omega \omega_\text{m}/Q)]^{-1}$), yields the GW strain sensitivity shown in Figs.~\ref{fig:FullPSD_1734Hz_5195Hz}d and e. The $[1,-1]$ and $[2,-2]$ modes achieve on-resonance sensitivities of $\sqrt{S_{hh}} = \unit[2 \times 10^{-18}]{Hz^{-1/2}}$ and $\sqrt{S_{hh}} = \unit[8 \times 10^{-19}]{Hz^{-1/2}}$, respectively.  These GW strain sensitivities are approximately $10^5$ times worse than advanced LIGO \cite{Martynov2016}, although with a detector that is approximately $10^5$ times shorter.  This suggests that with improvements, especially to the mechanical $Q$, our compact prototype represents a viable candidate for observation of GW sources.

\section{Conclusions}

The implementation of a re-entrant microwave cavity as a parametric transducer for acoustic modes in superfluid helium resulted in an enhanced coupling strength of $\unit[1.78]{GHz/bar}$, equivalent to a single-photon single-phonon coupling rate of up to $|g_0| = \unit[2.8 \times 10^{-5}]{Hz}$. This readout sensitivity is required to resolve the thermomechanical fluctuations of the helium normal modes, which ultimately limit the sensitivity of the detector. On resonance, with a $\approx \unit[4]{g}$ mass of helium and GW cross section of $\approx \unit[10]{cm^2}$, this detector achieves a sensitivity of $\unit[\approx 10^{-18}]{Hz^{-1/2}}$ for acoustic modes that couple to a quadrupolar GW strain.  It is also worth nothing that an array of such compact superfluid helium GW detectors may provide similar sensitivity to one detector the size of the array itself, while also allowing for coincidence analysis \cite{Fomalont1974,Napier1983,Bassan1996,Goryachev2014,Ratzel2019}.

A crucial quantity limiting the sensitivity of the detector is the mechanical quality factor, which saturated to around $10^6$ at millikelvin temperatures. The dominant loss mechanism was likely the mechanical $Q$ of the membrane, which could be improved in future experiments through careful treatment of high-$Q$ materials such as niobium \cite{Paik1976,Veitch1987}. 

Even though the minimum sensitivity is only achievable within a small detection bandwidth of less than a hertz, we demonstrated a tunability of the acoustic mode frequencies of $\unit[57.3]{Hz/bar}$ and $\unit[173]{Hz/bar}$ for the $[1,-1]$ and $[2,-2]$ mode, respectively. One could increase the pressure up to the melting curve of $^4$He at $\approx \unit[25]{bar}$ with a roughly linear increase of the sound speed \cite{Brooks1977,Abraham1970}, corresponding to a frequency tunability of $\approx \unit[50]{\%}$, as shown by the dashed lines in Fig.~\ref{fig:GWparameterspacePlot}. This would facilitate a broadband search for continuous GWs as well as enable the frequency corrections necessary to compensate for Doppler shifts caused by earth's motion \cite{Schutz1989}. We note that while we did not search for acoustic modes above 7 kHz in the current experiment, such higher order modes could also couple to GWs, although with diminishing $A_g$, as shown in Table \ref{tab:ModeParameters}.  As a result of the competing effects of frequency and cross section in the force spectrum, Eqn.~\eqref{Eq:GWForceStrainRelation}, these higher order modes would have relatively constant GW strain sensitives, as shown in Fig.~\ref{fig:GWparameterspacePlot}.

Finally, we note that although this detector is designed to search for GWs, it also serves as a prototype dark matter (DM) detector. Scalar ultralight dark matter can produce an isotropic strain signal that is qualitatively similar to a GW. For this reason it has been proposed to use resonant-mass GW detectors to search for dark matter \cite{Arvanitaki2016}, and data from some existing GW-detectors has even been reanalyzed to provide constraints on a scalar DM-induced strain \cite{Branca2017,Vermeulen2021}. A prototype DM detector with similar transduction but larger effective mass and mechanical quality factor is under development to more efficiently probe greater regions of unexplored DM parameter space. As a specific example, performing the sensitivity analysis detailed in Ref.~\cite{Manley2020}, we find that a cylindrical detector with 2 cm radius, 13 cm length, and $Q=10^6$, operating at 20 mK, can achieve a DM strain sensitivity of $\sqrt{S_{hh}} \approx \unit[2 \times 10^{-19}]{Hz^{-1/2}}$. This would be sufficient for the lowest symmetric mode at $\approx \unit[1800]{Hz}$ to beat the current constraints on scalar DM coupling within a few hours of integration time.

This work demonstrates key technical developments towards using macroscopic devices based on superfluid helium for the detection of weak forces. Along with searching for GWs, devices based on this architecture can also be used to probe couplings to the dark sector or gravitationally induced decoherence models.

\section*{Acknowledgments}

We would like to thank Qiyuan Hu for initial calculations and helpful discussions. This work was supported by the University of Alberta; the Natural Sciences and Engineering Research Council, Canada (Grants No. RGPIN-04523-16 and No. CREATE-495446-17); the Arthur B. McDonald Canadian Astroparticle Physics Research Institute through the support of the Canada First Research Excellence Fund, and US National Science Foundation Grant No. PHY-1912480.

\bibliography{references}{}

\end{document}